\documentclass[aps,prc,twocolumn,groupedaddress]{revtex4-2}
\vspace{5mm}

\usepackage{slashed}
\usepackage{graphicx}
\usepackage{braket}
\usepackage{amssymb}
\usepackage{mathtools}
\usepackage{bbold}
\usepackage{amssymb,latexsym}
\usepackage{amsmath,amsbsy,bbm}
\usepackage{multirow}
\usepackage[vcentermath]{youngtab}
\usepackage{nicefrac}
\usepackage[perpage]{footmisc}
\usepackage{wrapfig,lipsum,booktabs}
\usepackage{subcaption}
\usepackage{graphicx}
\usepackage{cjhebrew}
\usepackage{wrapfig,lipsum,booktabs}
\usepackage[justification=centering]{caption}
\usepackage{floatrow}
\usepackage[dvipsnames]{xcolor}

\newcommand{\simge}{\hspace*{0.2em}\raisebox{0.5ex}{$>$}
     \hspace{-0.8em}\raisebox{-0.3em}{$\sim$}\hspace*{0.2em}}
\newcommand{\simle}{\hspace*{0.2em}\raisebox{0.5ex}{$<$}
     \hspace{-0.8em}\raisebox{-0.3em}{$\sim$}\hspace*{0.2em}}

\usepackage[normalem]{ulem}

\begin{document}

\author{L. Contessi}\thanks{lorenzo@contessi.net}
\affiliation{Universit\'e Paris-Saclay, CNRS-IN2P3, IJCLab, 91405 Orsay, France}

\author{M. Sch\"{a}fer}
\email{m.schafer@ujf.cas.cz}
\affiliation{Nuclear Physics Institute of the Czech Academy of Sciences, \v{R}e\v{z} 25068, Czech Republic}

\author{U. van Kolck}
\affiliation{Universit\'e Paris-Saclay, CNRS-IN2P3, IJCLab, 91405 Orsay, France}
\affiliation{Department of Physics, University of Arizona, Tucson, Arizona 85721, USA}

\date{\today}

\title{Improved action for contact effective field theory}

\begin{abstract}
We present an improved action for renormalizable effective field theories (EFTs) of systems near the two-body unitarity limit. The ordering of EFT interactions is constrained, but not entirely fixed, by the renormalization group. 
The remaining freedom  
can be used to improve the theory's convergence, to simplify 
its applications,
and to connect it to
phenomenological models.
We exemplify the 
method on a contact theory 
applied to systems of up to five $^4$He atoms.
We solve the EFT at 
leading order including a subleading interaction that accounts for part of the two-body effective range. We show that the effects of such fake range 
can be compensated in perturbation theory at 
next-to-leading order, as long as 
the fake range is smaller or comparable to the experimental effective range.
These results open the possibility of using similar improved actions for other many-body systems.
\end{abstract}

\maketitle

\section{Introduction}

Systems near the two-body unitarity limit --- such as $^4$He atoms, where the two-body scattering length is about 20 times larger than the range of the interaction --- can be described systematically with effective field theories (EFTs) \cite{Hammer:2019poc}. At leading order (LO), in the EFT expansion for bosons, two- and three-body contact interactions reproduce Efimov physics \cite{Braaten:2004rn}. At next-to-leading order (NLO), the interaction range appears in the form of a two-derivative two-body contact interaction \cite{vanKolck:1999mw} and renormalization requires a four-body force \cite{Bazak:2018qnu}. Experiment and potential-model results for $^4$He can be approximated in a controlled way, with NLO improving \cite{Bazak:2018qnu} on an already good LO description of small \cite{Bazak:2016wxm} and large \cite{deleon2022equation} clusters.

EFTs incorporate universality in a few LO parameters and corrections to it at subleading orders. It can also be applied to fermions such as nucleons, and generalized to smaller distances by incorporating the longest-range interactions such as pion exchange in nuclear physics \cite{Hammer:2019poc}, the Van der Waals interaction between neutral 
atoms \cite{Odell:2021ryo}, and the induced-dipole interaction between a charged particle and a neutral atom \cite{Odell:2023gfd}. EFTs make no specific assumptions about the dynamics at distances much smaller than those of interest. They are particularly useful when the short-range dynamic is either unknown or hard to solve --- as is the case for 
Quantum Chromodynamics (QCD) in nuclear physics, where EFTs are now the mainstream
\cite{Hammer:2019poc}. 

However, current applications of EFT to many-body systems are not without problems. Most serious is the instability of multi-state fermion systems at LO. In contrast to experiments, the 6- \cite{Stetcu:2006ey}, 16- \cite{Contessi:2017rww,Bansal:2017pwn} and 40- \cite{Bansal:2017pwn} nucleon ground states are not stable at LO, even when pion interactions are included explicitly \cite{Yang:2020pgi}. Even though it works well for up to four nucleons \cite{Konig:2016utl}, an expansion around the unitarity limit is unlikely to produce stable states at LO \cite{Dawkins:2019vcr,Schafer:2020ivj}.

Stability in multi-component fermion systems can be a small effect, in the sense that the stable state is much closer to the nearest threshold than to the threshold for break up into individual constituents. For example, the $^{16}$O ground state is stable by only $\sim 15\%$ of its binding energy. There is indication that unstable states might exist not too far from threshold \cite{Contessi:2022vhn} and that they could be moved below threshold at higher orders in the EFT. Unfortunately, it is unclear how that can be accomplished in perturbation theory starting from 
the wavefunction of an
unstable state.

So far, the solution to this problem has been to ignore the deficiencies of LO and resum higher-order interactions into the exact solution of the many-body Schr\"odinger equation with truncated potentials \cite{Machleidt:2020vzm}. 
However, this approach produces unrenormalized amplitudes \cite{vanKolck:2020llt}: observables depend on the chosen, arbitrary regularization procedure. The problem stems from the increasingly singular nature of higher-order EFT interactions, which poses significant complications not seen for regular potentials. The simplest example is the resummation of range corrections, which can produce stable states \cite{Bansal:2017pwn} but whose renormalization violates \cite{Phillips:1996ae} the Wigner bound \cite{Wigner:1955zz}.
This represents a significant drawback 
since 
regulator dependence  
detracts from a clear hierarchy of interaction orders: results depend on the choice of parameters with arbitrarily many derivatives encoded in the specific regulator used. The supposed higher-order contributions are no longer small nor amenable to perturbation theory.
Connection with the underlying theory is 
obscured and {\it a priori} error control is
lost \cite{vanKolck:2020llt}. An alternative scheme for effective-range corrections, which might offer a chance to retain good renormalization properties even when the range corrections are partially resummed, has been proposed in Ref.~\cite{Ebert:2021ghj}. 

The root of an EFT's model independence is the renormalization group (RG), which requires that sufficiently many interactions be present at each order to remove essential dependence on the arbitrary regulator. At any order, the remaining cutoff dependence must be small, in the sense that it can be reduced arbitrarily with increasing momentum cutoff. Once this is achieved, cutoff errors are no larger than uncertainties arising from the truncation of the Hamiltonian, as long as the cutoff is taken beyond the breakdown scale of the EFT. 
Small contributions can always be resummed at a given order,
as long as RG invariance is preserved.
This is implicitly done when using data (which contain all orders in the EFT) to determine interaction strengths. One possibility to improve the efficiency of EFT in many-body systems is to determine interaction strengths from the corresponding data rather than few-body data.  
This approach is phenomenologically fruitful for unrenormalized interactions that include pions \cite{Ekstrom:2015rta}, but it precludes a systematic study of the possible breakdown of the EFT as density increases with particle number. 

We follow a different approach, inspired by the improved actions of lattice quantum field theory \cite{Symanzik:1983dc}. There, irrelevant interactions are added to relevant interactions to speed up convergence with a decreasing lattice spacing. Here we include at LO a subleading interaction that 
simplifies numerical calculations and thus may
bypass the LO instability problem. 
The clue to what interaction to include is offered by existing calculations (for example, Ref. \cite{Contessi:2017rww}), where stability is lost only at high momentum cutoffs.  
Since a finite momentum cutoff amounts to a finite minimum distance, we modify the LO interaction to account for a fake, or auxiliary, interaction range. We vary the fake range within the limits of the LO error, as estimated by standard power counting for contact EFTs. We show explicitly that at NLO its effects can be removed in perturbation theory. The fake range is thus not a physical parameter and power counting is preserved. 

Operationally, our approach brings the LO of the EFT close to models 
employed abundantly in nuclear and atomic physics, where a two-body potential of finite range is used, sometimes in conjunction with three-body potentials --- whose importance is justified by EFT renormalization only when the two-body potential has zero range \cite{Bedaque:1998kg,Bedaque:1998km}. However, in these models typically the potential, which is treated exactly, has its range fitted to data --- for one of many examples, see Ref. \cite{Kievsky:2021ghz}. In the EFT the physical two-body effective range remains a perturbative, NLO effect. The introduction of the fake range at LO is merely a way to account for small effects while preserving RG invariance and the hierarchy of interactions. The fake range slightly changes the position of shallow poles that already exist at LO and enables perturbation theory with respect to this new location. 

We demonstrate our method using a contact EFT for resonant bosonic systems,
which we present in Sec. \ref{sec:theory}. Up to four particles, this EFT is formally identical to Pionless EFT \cite{Hammer:2019poc} in the limit of Wigner's SU(4) symmetry,  
but it is simpler to benchmark against other theoretical approaches. Our results are given in Sec. \ref{sec:results}, where we take atomic $^4$He as a specific example because it exhibits
a clearer separation of scales than the nuclear problem. 
Details are relegated to Apps. \ref{App:A} and \ref{sec:tables}. 
The implications of this 
method to the logic behind EFT calculations
and its relation to  
phenomenological approaches are discussed in Sec. \ref{sec:discussion}.

\section{Theory}
\label{sec:theory} 

Contact EFT is a tool for describing nonrelativistic quantum systems where the typical many-body momentum is small compared to the inverse of the interaction range $R$ \cite{Hammer:2019poc}.  
In this theory, interparticle interactions are expanded in contact operators and derivatives, which require 
regularization and renormalization. 
We work in coordinate space where the position of particle $i$ is denoted by $\vec{r}_i$ and relative positions, by $\vec{r}_{ij}$. 
We consider a cutoff regularization with a momentum cutoff $\Lambda$, which replaces a Dirac delta function by 
a smeared function $\delta_{\Lambda}(\vec{r}_{ij})$.  

When the two-body scattering length $a_2$ dominates over other scales, such as the two-body effective range $r_2$, $|a_2|\gg |r_2|$, a subset of interactions needs to be treated nonperturbatively. 
The LO potential to be iterated to all orders is \cite{Hammer:2019poc}
\begin{equation}
    {\mathcal{V}}^{(0)}=V^{(0)}_{\text{2B}}+V^{(0)}_{\text{3B}},
    \label{VLO}
\end{equation}
in terms of two- and three-body interactions 
\begin{align}
    V^{(0)}_{\text{2B}}(\{\vec{r}_i\};\Lambda) 
    &= C_{0}^{(0)}(\Lambda) \sum_{i<j} \delta_\Lambda(\vec{r}_{ij}),
    \label{eq:EFTprime}\\
    V^{(0)}_{\text{3B}}(\{\vec{r}_i\};\Lambda)
    &= D_{0}^{(0)}(\Lambda) \sum_{i<j<k} \sum_{cyc} \delta_\Lambda(\vec{r}_{ij}) \delta_\Lambda(\vec{r}_{ik}).
    \label{eq:EFT}
\end{align}
The two-body interaction $V^{(0)}_{\text{2B}}$ guarantees the presence of a shallow two-body state \cite{vanKolck:1999mw}, while the three-body interaction $V^{(0)}_{\text{3B}}$ makes the three-body system well defined 
and introduces \cite{Bedaque:1998kg,Bedaque:1998km} 
a momentum scale which can be taken as
\begin{equation}
Q_3 \equiv \sqrt{\frac{2 m B_3}{3}},
\label{Q3}
\end{equation}
where $B_3$ is the binding energy of the three-body ground state.
The LO low-energy constants (LECs) $C_{0}^{(0)}(\Lambda)$ and $D_{0}^{(0)}(\Lambda)$ are chosen to reproduce a two-body datum like the scattering length $a_2$ (large or infinite) and a three-body datum, such as $B_3$.
All other observables --- for example, the ground-state binding energy $B_N$ of the $N$-body system --- can be predicted. The potential must depend on the cutoff so that observables do not. 

Subleading orders improve perturbatively on LO. 
At NLO there is a correction due to the two-body effective range $r_2$ \cite{vanKolck:1999mw}, which requires for renormalization also a contact four-body force, introducing
a four-body scale.
\cite{Bazak:2018qnu}.
The NLO (non-local) potential therefore reads
\begin{equation}
    \mathcal{V}^{(1)} = V^{(1)}_{\text{2B}} 
    + V^{(1)}_{\text{3B}} + V^{(1)}_{\text{4B}}
    \label{eq:NLO}
\end{equation}
with
\begin{eqnarray}
    V_{\text{2B}}^{(1)}(\{\vec{r}_i\};\Lambda)
    &=&C_{0}^{(1)}(\Lambda)\sum_{i<j} \delta_\Lambda(\vec{r}_{ij})
    \label{eq:nonlocalNLOprime}
    \\
    &+&C_{2}^{(1)}(\Lambda)\sum_{i<j} \left(\delta_\Lambda(\vec{r}_{ij}) \overrightarrow{\nabla}^2_{ij} + \overleftarrow{\nabla}^2_{ij} \delta_\Lambda(\vec{r}_{ij})\right),
    \nonumber \\
    V_{\text{3B}}^{(1)}(\{\vec{r}_i\};\Lambda) 
    &=&D_{0}^{(1)}(\Lambda) \sum_{i<j<k} \sum_{cyc} \delta_\Lambda(\vec{r}_{ij}) \, \delta_\Lambda(\vec{r}_{ik}),
    \label{eq:nonlocalNLOdoubleprime}\\
    V_{\text{4B}}^{(1)}(\{\vec{r}_i\};\Lambda) 
    &=& E_{0}^{(1)}(\Lambda) \sum_{i<j<k<l}\,
    \prod_{[a\neq b]\,\in \,\{i,j,k,l\}}\delta_\Lambda(\vec{r}_{ab}).
    \label{eq:nonlocalNLO}
\end{eqnarray} 
The LEC $C_2^{(1)}(\Lambda)$ accounts for the effective range, while $C_0^{(1)}(\Lambda)$ is the NLO momentum-independent LEC that ensures another two-body observable, for example
$a_2$, takes its desired value.
The LEC $D_{0}^{(1)}(\Lambda)$ can likewise be used to ensure $B_3$
remains unchanged. Finally, $E_{0}^{(1)}(\Lambda)$ is fixed by a four-body datum, for example $B_4$.
All other observables are predicted with improved accuracy.

If $|a_2|Q_3 \sim 1$, the finite value of the scattering length must be accounted for at LO through a LEC whose running with $\Lambda$ is parametrized by $a_2$, $C_0^{(0)}(\Lambda)\equiv C_0(a_2, \Lambda)$. In this case, we can choose $C_0^{(1)}(\Lambda)$ so as to keep $a_2$ unchanged at NLO. 
When, instead, $|a_2|Q_3\gg 1$, effects induced by the finiteness of the scattering length are subleading and can be introduced in an additional expansion in powers of $(a_2 Q_3)^{-1}$ \cite{Konig:2016utl}. In this case, $C_0^{(0)}(\Lambda)=C_0(a_2=\infty, \Lambda)$ is a fixed function of $\Lambda$ and two-body unitarity is exact at LO. Moreover, more-body systems exhibit discrete scale invariance \cite{Bedaque:1998kg,Bedaque:1998km}. Depending on its magnitude, the finite value of the scattering length could enter through $C_0^{(1)}(\Lambda)$. However, in this case, we can still include the finite scattering length at LO without destroying renormalizability.
The choice of the expansion point $a_2$ 
is to a certain extent arbitrary and any large value of the scattering length can be used as long as the perturbative correction to the physical $a_2$ value remains small. This preserves the model independence of the EFT since different choices of expansion points are equivalent up to the truncation error of each order. In fact, one frequently fits at LO not $a_2$ but the binding energy of the two-body system, with the difference being mostly corrected at NLO. The choice is usually guided by practical considerations, for example whether one is interested predominantly in bound-state properties or scattering. 
The formulation of contact EFT which considers the finite scattering length at LO and a perturbative effective-range correction 
together with a four-body scale at NLO
has been successfully applied to 
few-body systems, both atomic $^4\text{He}$ \cite{Bazak:2018qnu} 
and nuclear \cite{Schafer:2022hzo,Bagnarol:2023vhn}.

Accounting at LO for a finite scattering length even when close to the unitarity limit is an example of a more general procedure, where a subset of higher-order interactions $\Delta V$ are resummed into leading order, 
\begin{equation}
\tilde{\mathcal{V}}^{(0)}=V^{(0)}_{\text{2B}}+V^{(0)}_{\text{3B}} 
   +\Delta V.
\end{equation}
This 
partial resummation of subleading orders 
only changes the 
parameters used as the starting point to expand the theory.
It would be entirely trivial were we not seeking a systematic expansion with order-by-order renormalization and corrections amenable to perturbation theory. As such, we are restricted to a 
$\Delta V$ which does not change LO results by more than the expected LO error, and 
which can be compensated at higher orders by an ${\cal O}(1)$ relative shift in their LECs. 
This ensures that the new EFT expansion is equivalent to the standard one 
within the order-by-order truncation error. Since the potential is not observable, we cannot in general check this property without examining the convergence of physical 
quantities such as binding energies.

In the case of near unitarity, the inclusion of the scattering length amounts to 
\begin{equation}
\Delta V(\{\vec{r}_i\};\Lambda) = \left(C_0(a_2, \Lambda)-C_0^{(0)}(\Lambda)\right)\sum_{i<j} \delta_\Lambda(\vec{r}_{ij}).
\end{equation}
Accounting for $a_2$ at LO can be compensated by changes in the higher-order LECs such as $C_0^{(1)}(\Lambda)$.
One might consider, similarly,
to expand the theory around a small effective-range value $r_2>0$. 
However, this cannot be done with a $\Delta V$ in the form of a two-derivative contact interaction
because of the Wigner bound \cite{Wigner:1955zz}.

This issue can be circumvented by taking for $\Delta V$ an interaction with 
a small but finite range $\tilde{R}$,
which implicitly resums parts of not only the two-body effective range but also the higher parameters in the effective-range expansion.
The most natural --- but by no means unique --- way of redefining the expansion point 
is to take $\Delta V$ to have the same form as the regularized LO interaction,
\begin{equation}
\Delta V(\{\vec{r}_i\};\Lambda)= \sum_{i<j} 
\left(\tilde{C}\, \delta_{\tilde{R}^{-1}}(\vec{r}_{ij}) 
- C_0^{(0)}(\Lambda)\, \delta_\Lambda(\vec{r}_{ij})\right).
\label{eq:aux2b}
\end{equation}
The parameter $\tilde{C}$ can be fitted to the scattering length $a_2$. 
If the range $\tilde{R}$ is sufficiently small, results from the entire potential in Eq.~\eqref{eq:aux2b} will not deviate from those of
Eq.~\eqref{eq:EFTprime} more than the contributions from subleading corrections. Since the main effect of the potential \eqref{eq:aux2b} is to induce a fake effective range $\tilde{r}_2\sim \tilde R$, at NLO we fit the 
corresponding LECs to the correct $a_2$ and $r_2$. The procedure is repeated at higher orders.

A similar improvement can be performed with the three-body force,
\begin{eqnarray}
\Delta V(\{\vec{r}_i\};\Lambda)&=& \sum_{i<j} 
\left(\tilde{C}\,\delta_{\tilde{R}^{-1}}(\vec{r}_{ij}) 
- C_0^{(0)}(\Lambda)\,\delta_\Lambda(\vec{r}_{ij})\right)
\nonumber\\
&+&\sum_{i<j<k}\,\sum_{cyc }  \left(\tilde{D}\,\delta_{\tilde{R}^{-1}}(\vec{r}_{ij}) \, \delta_{\tilde{R}^{-1}}(\vec{r}_{ik}) 
\right.\nonumber\\
&&\left.\qquad
- D_0^{(0)}(\Lambda)\,\delta_\Lambda(\vec{r}_{ij}) \, \delta_\Lambda(\vec{r}_{ik})\right).
\label{eq:aux3b}
\end{eqnarray}
Here $\tilde{D}$ still fixes $B_3$,
while the additional effects of $\tilde{R}$ are minimized at next-to-next-to-leading (N$^2$LO), when the two-derivative contact three-body force appears \cite{Ji:2012nj}. Again, we must take $\tilde{R}$ small enough to preserve EFT convergence.

Other improvements could be made, but the two just mentioned are most natural since they modify interactions already present at LO. Even with this restriction, different fake ranges could be chosen for the two- and three-body improvements, similar to what is done in Ref. \cite{Kievsky:2017mjq}. This approach could be advantageous as it introduces additional flexibility, but also greater complexity.
To exemplify how the freedom in the 
organization of the EFT impacts its results, we consider the two different redefinitions of the LO interaction in Eqs. \eqref{eq:aux2b} and \eqref{eq:aux3b}, respectively
\begin{eqnarray}
\tilde{\mathcal{V}}_{I}^{(0)} &=&
\tilde{C}\sum_{i<j} \delta_{\tilde{R}^{-1}}(\vec{r}_{ij}) 
\nonumber\\
&+&D_{0}^{(0)}(\Lambda)\sum_{i<j<k} \sum_{cyc} \delta_\Lambda(\vec{r}_{ij}) \,\delta_\Lambda(\vec{r}_{ik}),
\label{curlyVI}
\\
\tilde{\mathcal{V}}_{II}^{(0)} &=&
\tilde{C}\sum_{i<j} \delta_{\tilde{R}^{-1}}(\vec{r}_{ij}) 
\nonumber\\
&+&\tilde{D}\sum_{i<j<k} \sum_{cyc} \delta_{\tilde{R}^{-1}}(\vec{r}_{ij}) \,\delta_{\tilde{R}^{-1}}(\vec{r}_{ik}). 
\label{curlyVII}
\end{eqnarray} 
The redefined LO interaction is iterated to all orders. In the case of $\tilde{\mathcal{V}}_{II}^{(0)}$, results are finite and the cutoff $\Lambda$ enters only in NLO calculations. For $\tilde{\mathcal{V}}_{I}^{(0)}$ the cutoff $\Lambda$ is needed already at LO because of the three-body force. The NLO interaction, Eq.~(\ref{eq:NLO}), is included in first-order perturbation theory. The form of the NLO interaction 
is not modified by the auxiliary interaction $\Delta V$. However, the cutoff dependence of the 
NLO LECs changes.
The LECs of Eqs. \eqref{eq:nonlocalNLOprime}, \eqref{eq:nonlocalNLOdoubleprime},
and \eqref{eq:nonlocalNLO} are, therefore, replaced by
\begin{eqnarray}
    C_{0}^{(1)}(\Lambda) &\rightarrow{}& \tilde{C}_{0}^{(1)}(\Lambda,\tilde{R}),  \nonumber \\
    C_{2}^{(1)}(\Lambda) &\rightarrow{}& \tilde{C}_{2}^{(1)}(\Lambda,\tilde{R}), \nonumber \\
    D_{0}^{(1)}(\Lambda) &\rightarrow{}& \tilde{D}_{0}^{(1)}(\Lambda,\tilde{R}), \nonumber\\
     E_{0}^{(1)}(\Lambda) &\rightarrow{}& \tilde{E}_{0}^{(1)}(\Lambda,\tilde{R}).
\end{eqnarray}

The fact that $\Delta V$ is a subset of higher-order interactions automatically guarantees that a sufficiently small $\tilde{R}$ exists where the above procedure can be carried out. What the maximum $\tilde{R}$ is a question which we address in the following.

\section{Atomic $^4$H\lowercase{e} results}
\label{sec:results} 

We analyze $\tilde{\mathcal{V}}_{I}^{(0)}$ and $\tilde{\mathcal{V}}_{II}^{(0)}$ in systems ranging from two to five bosons. 
We focus on the behaviour of LO and NLO
with respect to the 
cutoff $\Lambda$ and the range $\tilde{R}$ of the auxiliary interaction. 
For convenience, we display the $\Lambda$ and $\tilde{R}^{-1}$ dependence in units of the three-body scale $Q_3$, Eq.~(\ref{Q3}).
To be definite, we 
employ a Gaussian smearing of the Dirac delta function,
\begin{equation}
\delta_\Lambda(\vec{r})=\left(\frac{\Lambda}{\sqrt{\pi}}\right)^3~\text{exp}\left(-\dfrac{\Lambda^2}{4}\vec{r}^{\, 2}\right).
\label{eq:regulator}
\end{equation}
The large-cutoff limit of an observable $\mathcal{O}$ can be obtained from a fit
\begin{equation}  
\mathcal{O}(\Lambda)=\mathcal{O}(\infty)+\frac{a}{\Lambda} +\frac{b}{\Lambda^2} 
\label{eq:LO_extrap}
\end{equation}
with fitting parameters $a$, $b$. The $\Lambda^{-1}$ term accounts for contributions that will be removed at the next order
once the cutoff exceeds the theory's breakdown scale, while the $\Lambda^{-2}$ term provides further stability to the fits for moderate cutoff values.

We test our theory in 
systems made out of $^4 \text{He}$ atoms, whose inverse mass is 
$(\hbar c)^2/m=12.119~{\rm K \cdot \AA^2}$.
As in previous EFT work \cite{Bazak:2016wxm,Bazak:2018qnu}, we constrain our theory with properties calculated \cite{doi:10.1063/1.481404,doi:10.1063/1.469978,PhysRevA.70.052711,PhysRevA.73.062717,Hiyama:2011ge} 
with either LM2M2 \cite{LM2M2:1991} or PCKLJS \cite{PCKLJS:2010} $^4\text{He}$-$^4\text{He}$ potential models:
The values listed in Table~\ref{tab:observables} are used here in lieu of experimental data, since only a few experimental values \cite{Grisenti:2000zz, Kunitski:2015qth, Zeller:2016mwo} --- also listed in Table \ref{tab:observables} --- are known. Experimental data could replace potential-model input once more of them are available. 

\begin{table}[bt]
    \centering
\begin{tabular}{l  r r r}
\hline \hline
     & ~~LM2M2&  ~~~~PCKLJS& ~~~~~~~~~~~~~~~~~Exp.  
     \\ \hline
 $a_2$ (\AA) & 100.23& 90.42(92)& \\  
 $r_2$ (\AA) & 7.326& 7.27&  \\
 [5pt]
 $B_2$ (mK)  &  1.3094& 1.6154&~~~$1.3^{+0.25}_{-0.19}$; 1.76(15)\\
 $B_3$ (mK)  & 126.50& 131.84& \\
 $B_3^*$ (mK) & 2.2779& 2.6502& \\
 $B_3^* - B_2$ (mK) & 0.9685& 1.0348& 0.98(2)\\
 $B_4$ (mK) & 559.22& 573.90& \\
 $B_5$ (mK) & 1306.7& -& \\
 \hline \hline
\end{tabular}
   \caption{Values for the two-body scattering parameters
   (scattering length $a_2$ and effective range $r_2$) and $N$-body binding energies (ground states $B_N$ and first-excited state $B_3^*$) of atomic $^4$He systems from two phenomenological potentials LM2M2 \cite{LM2M2:1991} and PCKLJS \cite{PCKLJS:2010}, as extracted from Refs. \cite{doi:10.1063/1.481404,doi:10.1063/1.469978,PhysRevA.70.052711,PhysRevA.73.062717,Hiyama:2011ge}.
   Known experimental values \cite{Grisenti:2000zz, Kunitski:2015qth, Zeller:2016mwo} are given for comparison. }
    \label{tab:observables}
\end{table}

The structure of these atomic clusters is tied to the three-body scale $Q_3\simeq 0.08$~\AA$^{-1}$. Since $|a_2| Q_3\simeq 8$,
we expect an expansion around the unitarity limit to converge. Nevertheless, for simplicity --- and to compare with previous calculations \cite{Bazak:2016wxm,Bazak:2018qnu} --- we improve our LO to reproduce the $^4\text{He}$-$^4\text{He}$ scattering length $a_2$, in addition to the ground-state binding energy of the $^4\text{He}$ trimer, $B_3$. 
At two-body level, the LO scattering fit is done using the Numerov algorithm. The three-, four-, and five-body systems are solved with the stochastic variational method (SVM) \cite{Suzuki:1998bn}.
We predict the binding energies of the dimer, $B_2$, trimer first-excited state, $B_3^*$, and of the tetramer and pentamer ground states, $B_4$ and $B_5$. 
At NLO, we employ the distorted-wave Born approximation.
We fit 
the 
effective range $r_2$ and the ground-state binding energy of the 
tetramer, $B_4$, 
and check for improvements in $B_2$, $B_3^*$ and $B_5$. 

Since we do not include the Van der Waals interaction explicitly, we would expect the expansion parameter to be ${\cal O} (r_{\rm vdW} Q)$, where $r_{\rm vdW}\sim 5$\AA~$\sim r_2$ and $Q$ is the typical particle momentum in the bound state. For dimer properties, where $Q\sim a_2^{-1}$, the errors of the expansion are expected 
to be 
$\sim r_2/a_2\simeq 10\%$ at LO and $\sim r_2^2/a_2^2\simeq 1\%$ at NLO. The errors for more-body states are likely larger but harder to estimate. Taking $Q\sim Q_3$, the only scale in the unitarity limit, the errors would be $\sim 40\%$
at LO and $\sim 15\%$
at NLO. These estimates include the power-counting argument that {\it (i)} the four-body force enters at NLO and thus should produce LO errors comparable to those of $r_2$; and {\it (ii)} the error at NLO should be given by the second Born approximation for NLO interactions as well as the first Born approximation for 
N$^2$LO interactions, such as 
a three-body momentum-dependent force.
Nevertheless, previous studies of Contact
EFT for $^4$He atomic systems \cite{Bazak:2016wxm,Bazak:2018qnu} suggest that these errors are overestimated, perhaps because $Q_3$ is an overestimate of the particle binding momentum.

In this section we show results of the theory fitted to LM2M2 quantities using a derivative NLO term, Eq.~(\ref{eq:nonlocalNLO}). The same calculations were performed also for a local, $r^2$ version defined in App.~\ref{App:A}.  
In App.~\ref{sec:tables} we summarize our numerical results calculated with derivative and $r^2$ NLO interactions for both LM2M2 and PCKLJS input.
Our main conclusions are independent of the specific input and form adopted for the NLO interaction.

\subsection{Leading order}

\begin{figure}[tb]
\centering 
  \includegraphics[width=\linewidth]{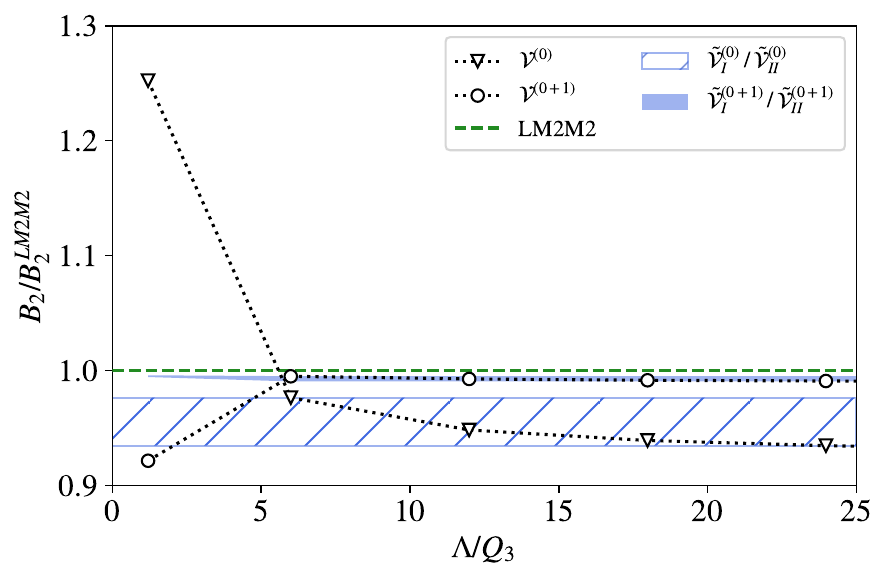}
  \caption{Calculated dimer binding energy normalized to the dimer binding energy obtained directly from the LM2M2 potential \cite{LM2M2:1991}, 
  $B_2/B_2^{\rm LM2M2}$, 
  as function of the cutoff $\Lambda$ in units of the three-body scale $Q_3$. 
  Results from the LO and NLO potentials, respectively $\mathcal{V}^{(0)}$ and $\mathcal{V}^{(0+1)}\equiv \mathcal{V}^{(0)}+\mathcal{V}^{(1)}$, are shown by 
  triangles and circles connected with dotted lines.
  At the two-body level, the improved potentials of type $I$ and $II$ are equivalent.
  The (blue) hatched and shaded bands 
  show the LO and NLO values from the improved potentials $\tilde{\mathcal{V}}^{(0)}_{I}$/$\tilde{\mathcal{V}}^{(0)}_{II}$ and $\tilde{\mathcal{V}}^{(0+1)}_{I}$/$\tilde{\mathcal{V}}^{(0+1)}_{II}$,
  respectively, when the fake range is in the range $6 \leq (Q_3 \tilde{R})^{-1} \leq 24$.   }    
  \label{fig:2bLONLO}
\end{figure}

When the LO interaction $\mathcal{V}^{(0)}$, Eq.~\eqref{VLO}, is fitted to the scattering length $a_2$, the resulting dimer binding energy $B_2$ depends on the momentum cutoff $\Lambda$ as shown in Fig. \ref{fig:2bLONLO}. In the zero-range limit,  
$B_2(\Lambda \rightarrow \infty ) = 1/m a_2^2$
and the ratio to the energy 
obtained directly from the LM2M2 potential, $B_2^{\rm LM2M2}$ (see Table~\ref{tab:observables}), is 
$B_2(\Lambda \rightarrow \infty )/B_2^{\rm LM2M2}\simeq 0.93$. 
As expected, results vary by less than 10\% once the cutoff exceeds the expected breakdown scale, $\Lambda/Q_3 \simge (r_{\rm vdW}Q_3)^{-1}\sim 3$.

Closely related are the results of the 
redefined LO potentials $\tilde{\mathcal{V}}^{(0)}_I$, Eq.~\eqref{curlyVI}, and $\tilde{\mathcal{V}}^{(0)}_{II}$, Eq.~\eqref{curlyVII}, 
which are equivalent at the two-body level.
Since their two-body parts depend solely on the size of the fake range $\tilde{R}$, no $\Lambda$ dependence is introduced.
In Fig.~\ref{fig:2bLONLO} the resulting dimer energies from $\tilde{\mathcal{V}}^{(0)}_I$ or $\tilde{\mathcal{V}}^{(0)}_{II}$ would show as horizontal lines that intersect the calculated $\mathcal{V}^{(0)}(\Lambda)$ energies 
where the inverse fake range reaches the cutoff, 
$\tilde{\mathcal{V}}^{(0)}_{I}(\tilde{R}^{-1} = \Lambda) = \tilde{\mathcal{V}}^{(0)}_{II}(\tilde{R}^{-1} = \Lambda) = \mathcal{V}^{(0)}(\Lambda)$. For clarity, we represent the results for $6 \leq (Q_3\tilde{R})^{-1}\leq 24$ as a 
horizontal band.
The band 
thus matches the residual cutoff dependence of $\mathcal{V}^{(0)}$.
For decreasing fake range, $B_2/B_2^{\rm LM2M2}$ attains the $B_2(\Lambda \rightarrow \infty )/B_2^{\rm LM2M2}$ value. As the fake range increases, the ratio stays well within the expected $\sim 10\%$ truncation error and reaches
$1$
for $\tilde{R}_c^{-1}\simeq 4.8 \, Q_3$, where the physical effective range is roughly reproduced already at the LO. Allowing even larger fake ranges leads to a deterioration of our results --- this region will be  
addressed 
in Sec. \ref{sec:maxfake}. The 
values of $B_2/B_2^{\rm LM2M2}$ at various fake ranges can be found in Table~\ref{tab:extrapolated_results2} of App.~\ref{sec:tables}.

\begin{figure*}[t]
\begin{tabular}{cc}
  \includegraphics[width=0.5\linewidth]{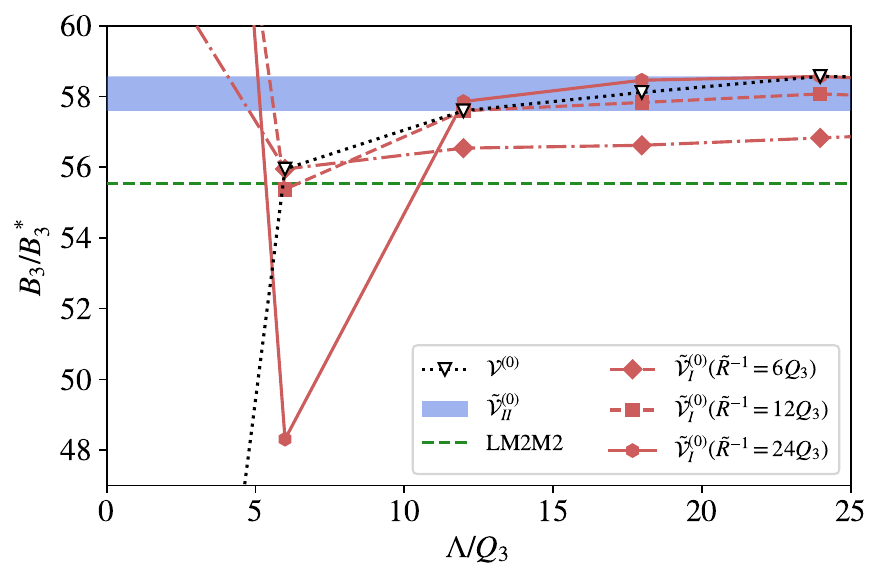}
&
  \includegraphics[width=0.5\linewidth]{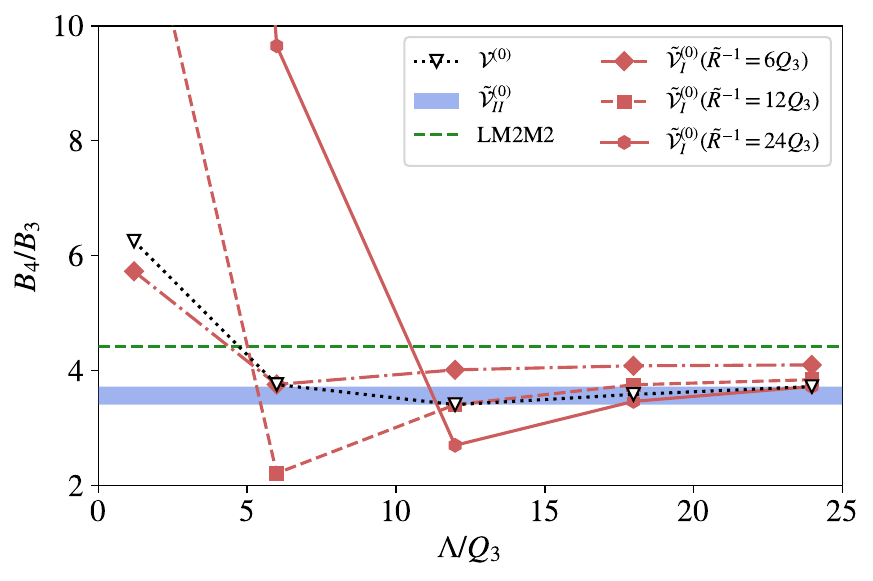}\\
\end{tabular}
  \caption{Binding energies for the excited trimer $B_3^*$ (left panel) and ground tetramer $B_4$ (right panel) at LO, normalized to the ground-trimer binding energy $B_3$, as function of the cutoff $\Lambda$ in units of the three-body scale $Q_3$. 
  Results for the LO potential $\mathcal{V}^{(0)}$ are shown by
  triangles connected with a dotted line.
  Results for the improved LO potential $\tilde{\mathcal{V}}^{(0)}_{I}$ 
  are given in red with different symbols for different 
  inverse fake-range values: $\tilde{R}^{-1} = 6\, Q_3$~(diamonds connected by a dash-dotted line), $\tilde{R}^{-1} = 12\, Q_3$~(squares, dashed line), and $\tilde{R}^{-1} = 24\, Q_3$~(circles, solid line). 
  The blue shaded band shows the LO values from $\tilde{\mathcal{V}}^{(0)}_{II}$ when the fake range is in the range $6 \leq (Q_3 \tilde{R})^{-1} \leq 24$. 
  The (green) horizontal dashed lines mark the result 
  of the full LM2M2 potential \cite{Hiyama:2011ge}.}
  \label{fig:LO1}
\end{figure*}

\begin{figure}[t]
\centering 
  \includegraphics[width=\linewidth]{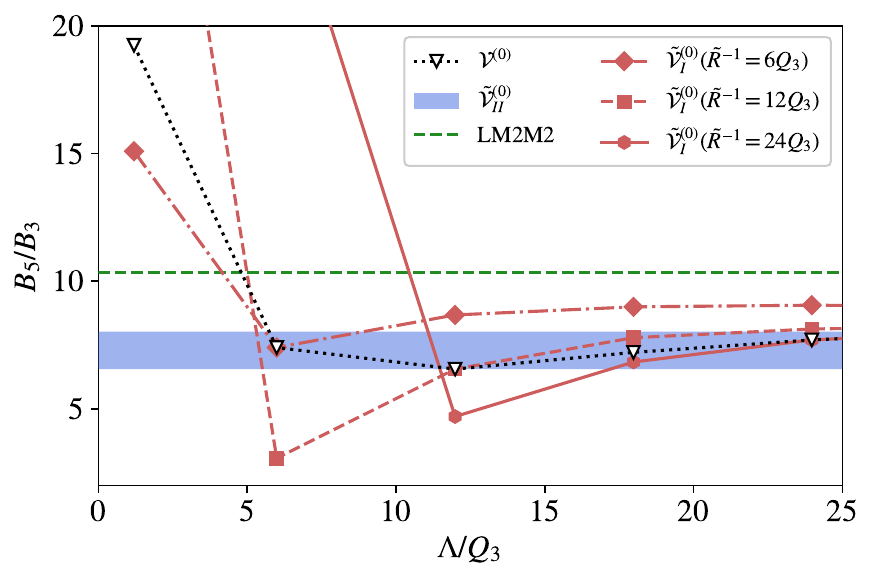}
  \caption{Same as Fig.~\ref{fig:LO1} but for the ground-pentamer binding energy $B_5$. The LM2M2 result is
  from Ref.~\cite{doi:10.1063/1.481404}
  }
  \label{fig:LO2}
\end{figure}

In Fig.~\ref{fig:LO1}, we present the LO predictions for the $B_3/B_3^*$ (left panel) and $B_4/B_3$ (right panel) binding energy ratios as a function of the increasing cutoff $\Lambda$. Predictions for $B_5/B_3$ are given in
Fig.~\ref{fig:LO2}. 
The results are for the improved LO  
potentials $\tilde{\mathcal{V}}^{(0)}_I$ and $\tilde{\mathcal{V}}^{(0)}_{II}$ 
with a few representative values of the inverse fake range in the range 
$6\le (Q_3 \tilde{R})^{-1} \le 24$.
We compare them to the standard LO potential $\mathcal{V}^{(0)}$.

The ratios obtained with 
$\mathcal{V}^{(0)}$ converge with $\Lambda$, and using the extrapolation function \eqref{eq:LO_extrap} we obtain 
$B_3/B_3^*(\Lambda\to\infty)=59.089$,
$B_4(\Lambda\to\infty)/B_3 = 4.150(6)$, 
and $B_5(\Lambda\to\infty)/B_3 = 9.27(5)$,
which are consistent with 
Refs.~\cite{Bazak:2016wxm,Bazak:2018qnu}. Here, the number in the parenthesis represents the numerical uncertainty, if non-negligible, which should be further inflated by the expected 
LO theoretical error. 

Because of the three-body force, the calculated $\tilde{\mathcal{V}}^{(0)}_I$ energies depend on $\Lambda$.
While $\mathcal{V}^{(0)}$ results show the expected residual cutoff dependence $\propto \Lambda^{-1}$, $\tilde{\mathcal{V}}^{(0)}_I$ energies appear, based on the limited amount of points available, to be compatible with a $\Lambda^{-2}$ behavior at large $\Lambda$.
This would imply that the $\Lambda^{-1}$ dependence of LO results
--- which with an unimproved action is compensated by NLO corrections ---
is removed by the fake range.
This is not unexpected for
two- and three-body systems, where 
improvement will include errors of order $\tilde{R}$ instead.
Since the four-body force is driven by the two-body range,
we cannot conclude that
the absence of $\Lambda^{-1}$ dependence in higher-body systems implies
that the four-body force is not strictly required for NLO renormalizability. 
This conclusion would contradict evidence from previous work~\cite{Bazak:2018qnu,Schafer:2022hzo}.

For $6 \leq (Q_3 \tilde{R})^{-1} \leq 24$ the calculated $\tilde{\mathcal{V}}^{(0)}_I$ binding energies converge 
at sufficiently large $\Lambda$ to values that are within the LO error of the full LM2M2 result. We observe that all $\tilde{\mathcal{V}}^{(0)}_I$ results get closer to the LM2M2 values the larger $\Lambda$ and the fake range $\tilde{R}$ are.
Cutoff-extrapolated values for various fake ranges are given in Table
\ref{tab:extrapolated_results_2range} of App. \ref{sec:tables}.

On the other hand, when the cutoff is smaller than the inverse fake range but larger than the supposed breakdown scale of the theory, we see an enhancement of the cutoff dependence in the results. 
This phenomenon is best illustrated by the behavior of $\tilde{\mathcal{V}}^{(0)}_{I}(\tilde{R}^{-1}=24Q_3)$ in Fig.~\ref{fig:LO2}, where the calculated pentamer $B_5/B_3$ ratios at $\Lambda=6\,Q_3$ and $12\,Q_3$ are very different from
the same quantities with the unimproved action.
For $\Lambda$ larger than the breakdown scale of the theory,
the convergence pattern sets in only for $\Lambda \simge \tilde{R}^{-1}$.
Since $\tilde{R}^{-1}$ functions as a two-body cutoff, this effect is in agreement with what was found in Ref.~\cite{Rotureau:2010uz}.

In contrast to $\tilde{\mathcal{V}}^{(0)}_{I}$, 
the 
improved $\tilde{\mathcal{V}}^{(0)}_{II}$ interaction depends solely on the fake range. 
As for the dimer, we depict the $\Lambda$-independent $\tilde{\mathcal{V}}^{(0)}_{II}$ results as straight horizontal bands which
represent 
the range $6 \leq (Q_3\tilde{R})^{-1}\leq 24$. Again, the corresponding bands match the residual cutoff dependence of 
the $\mathcal{V}^{(0)}(\Lambda)$ results. 
Tables \ref{tab:extrapolated_results34} and \ref{tab:extrapolated_results5}
of App. \ref{sec:tables} give 
the energy ratios at various fake ranges.
Note that $\mathcal{V}^{(0)}$ results converge from below, but to a larger value than the exact result only for the trimer excited state. Thus, only in this case does $\mathcal{V}^{(0)}_{II}$ give results closer to exact than the asymptotic $\mathcal{V}^{(0)}$ value. In any case, the width of the $\mathcal{V}^{(0)}_{II}$ band is not larger than the difference between $\mathcal{V}^{(0)}$ and exact results.

Partially accounting for the physical two-body effective range through the fake range ($\tilde{\mathcal{V}}^{(0)}_{I}$) brings LO results closer to the ``exact'' results. Incorporating this fake range in the three-body force as well ($\mathcal{V}^{(0)}_{II}$) does not always accomplish the same, perhaps reflecting the fact that the range of the three-body force is not fixed by that of the two-body force. 
On the other hand, the different ranges $\tilde{R}$ and $\Lambda^{-1}$ of, respectively, two- and three-body interactions in $\tilde{\mathcal{V}}^{(0)}_{I}$ should be selected with caution, as it is often required that $\Lambda \simge \tilde{R}^{-1}$. 
Moreover, 
$\tilde{\mathcal{V}}^{(0)}_{II}$ decreases significantly the number of few-body calculations to be performed at this order since it does not require independent variation of both ranges. 

In either case, the range of results is comparable to the difference between standard LO ($\mathcal{V}^{(0)}$) and exact model results. We examine next the extent to which fake range effects can be removed perturbatively at NLO.

\subsection{Next-to-leading order}

At NLO, we study 
binding energies obtained with the improved potentials $\tilde{\mathcal{V}}_I^{(0 + 1)}\equiv \tilde{\mathcal{V}}^{(0)}_{I}+\mathcal{V}^{(1)}$ and $\tilde{\mathcal{V}}_{II}^{(0 + 1)}\equiv \tilde{\mathcal{V}}^{(0)}_{II}+\mathcal{V}^{(1)}$ and how they compare with the standard potential 
$\mathcal{V}^{(0+1)}\equiv \mathcal{V}^{(0)}+\mathcal{V}^{(1)}$.

The LM2M2 effective range $r_2$, given in Table~\ref{tab:observables}, is fitted at NLO. The cutoff convergence of $B_2$, obtained from $\mathcal{V}^{(0 + 1)}$, is seen in Fig.~\ref{fig:2bLONLO} to rapidly reach 
the exact result within 1\%, as expected. As a consequence, the width of the horizontal band for the improved potential at NLO, $\tilde{\mathcal{V}}^{(0+1)}_{I}(\tilde{R}^{-1} = \Lambda) = \tilde{\mathcal{V}}^{(0+1)}_{II}(\tilde{R}^{-1} = \Lambda) = \mathcal{V}^{(0+1)}(\Lambda)$
with $6 \leq (Q_3\tilde{R})^{-1}\leq 24$, is narrow and
much smaller than the change in standard results from LO to NLO. 
As before, as long as the fake range remains smaller than the physical effective range, improved-potential results preserve the accuracy of the unimproved results. 

\begin{figure*}[t]
\begin{tabular}{cc}
  \includegraphics[width=0.5\linewidth]{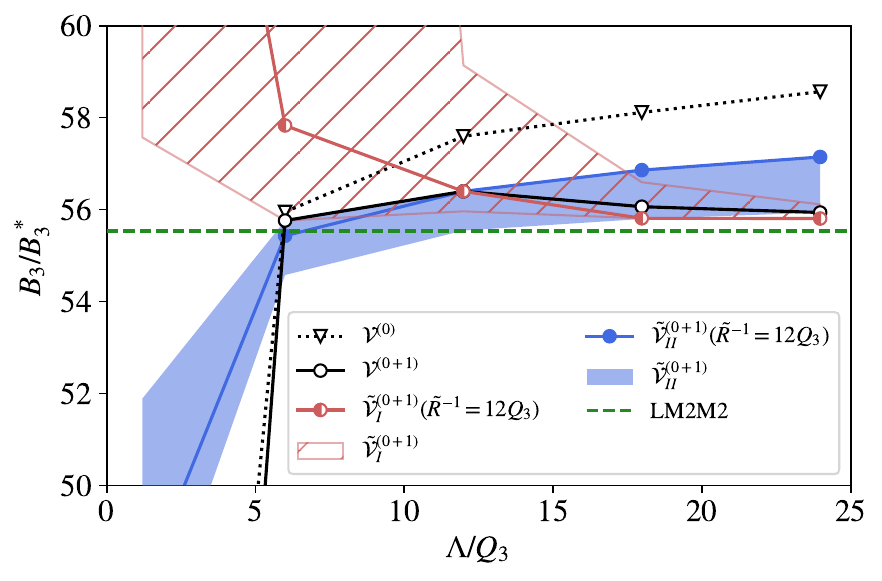}
&
  \includegraphics[width=0.5\linewidth]{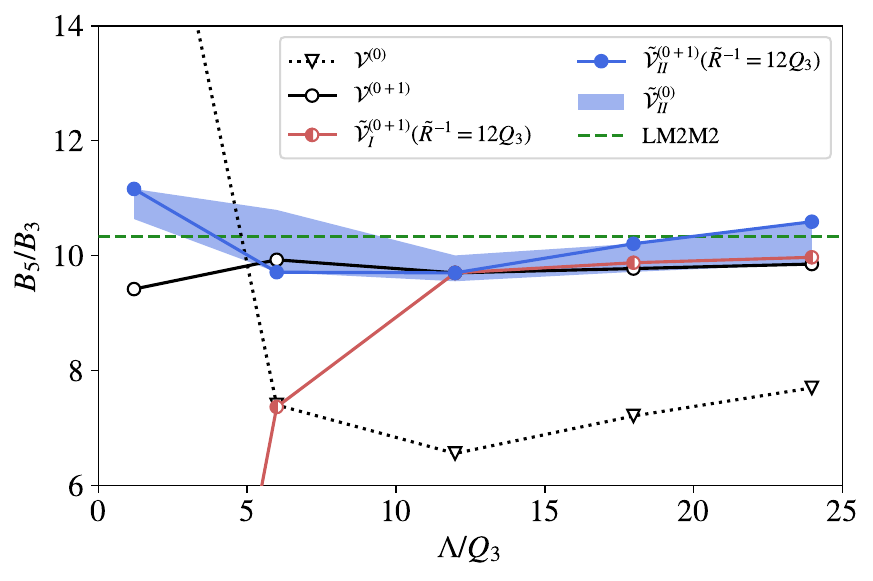}\\
    \end{tabular}
  \caption{Binding energies for the excited trimer $B_3^*$ (left panel) and ground pentamer $B_5$ (right panel) up to NLO, normalized to the ground-trimer binding energy $B_3$, as function of the cutoff $\Lambda$ in units of the three-body scale $Q_3$. 
  Results for the LO and NLO potentials $\mathcal{V}^{(0)}$ and $\mathcal{V}^{(0+1)}$ are shown by, respectively
  empty triangles connected with a dotted line and empty circles connected with a solid line.
  Results for the improved NLO potentials $\tilde{\mathcal{V}}^{(0+1)}_{I}$ and $\tilde{\mathcal{V}}^{(0+1)}_{II}$
  at $\tilde{R}^{-1} =12 \, Q_3$ are given by, respectively, (red) half-filled circles connected by a (red) solid line and (blue) solid circles connected by a (blue) solid line.
  The (red) hatched area and the (blue) shaded band indicate the spread of $\mathcal{V}^{(0+1)}_I$ and $\mathcal{V}^{(0+1)}_{II}$ results, respectively, for fake-range values in the range $6\leq (Q_3\tilde{R})^{-1} \leq 24$. (For the pentamer, we show $\mathcal{V}^{(0+1)}_{I}$ results only for one fake-range value, as obtaining accurate results for the entire range is numerically challenging.)
  The (green) horizontal dashed lines mark the result of the full LM2M2 potential
  \cite{Hiyama:2011ge, doi:10.1063/1.481404}.
  }
  \label{fig:NLO}
\end{figure*}

The ratio $B_4/B_3$ is used to constrain the NLO four-body interaction, thus reproducing the LM2M2 line on the right panel of 
Fig.~\ref{fig:LO1}.
The calculated $B_3/B_3^*$ 
and $B_5/B_3$ 
binding-energy ratios are shown as a function of the momentum cutoff $\Lambda$ in Fig.~\ref{fig:NLO}. 
As a benchmark, we extrapolate our $\mathcal{V}^{(0+1)}$ trimer and pentamer results for $\Lambda \rightarrow \infty$ using Eq.~(\ref{eq:LO_extrap}). We obtain $B_3/B_3^*(\Lambda\to\infty) = 55.803$ and $B_5(\Lambda\to\infty)/B_3 = 10.17(8)$. The extrapolated pentamer ratio agrees with the earlier result of Ref.~\cite{Bazak:2018qnu}, which was obtained from
the same $\mathcal{V}^{(0+1)}$ potential
renormalized to 
LM2M2 quantities.
For the trimer, the ratio obtained here slightly differs from
Ref.~\cite{Bazak:2018qnu}, which employed PCKLJS quantities as reference. 
Even better agreement is found using the PCKLJS setting, as can be seen in Table~\ref{tab:extrapolated_results34} of App.~\ref{sec:tables}. As noted in Ref.~\cite{Bazak:2018qnu}, the change of only $\sim 10\%$ compared to LO suggests the expansion parameter is smaller than expected, with asymptotic NLO values differing from exact ones by at most a few percent.

We have found our $\tilde{\mathcal{V}}^{(0+1)}_{I}$ NLO calculations 
to be rather sensitive to the 
accuracy in the description of the LO 
wavefunction. 
The sensitivity is especially pronounced 
when the LO two- and three-body potentials 
have very different ranges $\tilde{R}$ and $\Lambda^{-1}$.
As a result, large cancellations occur in various potential channels, which in turn lead to relatively slow convergence in our few-body calculations. While for trimer and tetramer we were able to 
obtain accurate
wavefunctions, we encountered challenges
with the pentamer ground state. Consequently, for $\tilde{\mathcal{V}}^{(0+1)}_{I}$ we depict in 
Fig.~\ref{fig:NLO} the $B_3/B_3^*$ ratio for a range of fake-range values, $6 \leq (Q_3\tilde{R})^{-1} \leq 24$, but
$B_5/B_3$ 
only 
for $\tilde{R}^{-1}=12\, Q_3$. For clarity, pentamer results for 
smaller $\tilde{R}^{-1}$ are not displayed in the figure,
but the corresponding $\Lambda \rightarrow \infty$ values are given in App.~\ref{sec:tables}. Results for 
larger $\tilde{R}^{-1}$ 
are at the moment beyond the reach of our few-body method. 

The resulting $\tilde{\mathcal{V}}^{(0+1)}_{I}$ values for the $B_3/B_3^*$ and $B_5/B_3$  
ratios follow the 
pattern 
observed when using solely the LO $\tilde{\mathcal{V}}^{(0)}_{I}$ part:
rather fast stabilization once the momentum cutoff $\Lambda$ is large enough to suppress regulator-induced, and thus arbitrary, effects such as
a finite range $\propto \Lambda^{-1}$ in the 
three-body force. 
Comparison with Fig.~\ref{fig:LO1} shows that 
the perturbative 
NLO correction $\tilde{\mathcal{V}}^{(1)}$ 
shifts the 
$B_3/B_3^*$ and $B_5/B_3$ ratios closer to the LM2M2 values. At the same time, the residual $\Lambda$ dependence decreases with respect to the LO results. 

The $\tilde{\mathcal{V}}^{(0+1)}_{II}$ type of improved EFT potential introduces $\Lambda$ dependence solely through the perturbative inclusion of $\mathcal{V}^{(1)}$ at NLO. The variation of our $\tilde{\mathcal{V}}^{(0+1)}_{II}$ results 
in the range $6 \leq (Q_3 \tilde{R})^{-1} \leq 24$ is shown for the trimer as well as the pentamer in Fig.~\ref{fig:NLO}. 
For any given $\tilde{R}^{-1}$ in this interval, the resulting $B_3/B_3^*$ and $B_5/B_3$ ratios stabilize with the increasing cutoff. 
The corresponding quantities, extrapolated to $\Lambda \rightarrow \infty$, are listed in Tables~\ref{tab:extrapolated_results34} and \ref{tab:extrapolated_results5} of App.~\ref{sec:tables}. 
There is a considerably milder residual cutoff dependence over the larger cutoff region than in the $\tilde{\mathcal{V}}^{(0+1)}_{I}$ case. This is likely linked to the same fake ranges being employed in both two- and three-body potentials 
of the LO $\tilde{\mathcal{V}}^{(0)}_{II}$ interaction. 

Computationally, there is a significant difference between the $\tilde{\mathcal{V}}^{(0+1)}_{I}$ and $\tilde{\mathcal{V}}^{(0+1)}_{II}$ few-body calculations, since for $\tilde{\mathcal{V}}^{(0)}_{II}$
the Schr\"{o}dinger equation needs to be solved only once for each fake range. This is especially beneficial once 
many $\tilde{R}$ values need to be considered. Furthermore, in the $\tilde{\mathcal{V}}^{(0+1)}_{II}$ case we do not observe 
severe numerical problems. This is likely related to the same $\tilde{R}$ values for
the two- and three-body LO 
potentials.

Despite the computational differences, the NLO results are qualitatively similar for both types of improvements we consider here. 
For all trimer NLO results we notice relatively fast convergence with both increasing $\Lambda$ value and EFT order. This can be understood by the shallow nature of the trimer excited state where the corresponding small typical momentum induces only a minor contribution from the further subleading corrections. 
The inclusion of the NLO terms yields a $B_3/B_3^*$ ratio which is already close to the ``exact'' value calculated directly with the LM2M2 potential. Even for the pentamer the cutoff dependence is mild once the cutoff is somewhat higher than for the trimer, probably as a consequence of a higher characteristic binding momentum. NLO results also improve significantly if compared with the exact result.  

We observe that the trimer $\tilde{\mathcal{V}}^{(0+1)}_{II}$ band is relatively wide even at large cutoff values. 
For this improved potential, the $\tilde{R}^{-1} = 12 \, Q_3$ results tend to differ the most from both the standard EFT potential and the ``exact'' LM2M2 value.
Excluding the $B_3/B_3^*$ ratios calculated for fake ranges close to $\tilde{R}^{-1} = 12 \, Q_3$, the $\tilde{\mathcal{V}}^{(0+1)}_{II}$ band shrinks significantly, making it more similar at large $\Lambda$ to the band calculated with $\tilde{\mathcal{V}}^{(0+1)}_{I}$. 
In $\tilde{\mathcal{V}}^{(0+1)}_{I}$ the three-body fake range is absent.
Therefore,
we attribute this large width primarily to a still substantial influence of the subleading three-body contributions induced around $\tilde{R}^{-1} = 12 \, Q_3$ by accidental cancellations.
The same statements hold for the pentamer, but the effect seems to be quenched by the presence of the four-body force.

Most importantly, at large cutoff, NLO results for the two types of improvement differ only by a few percent. They are very close to the non-improved NLO results, demonstrating that, at the NLO level, the choice of LO interaction was irrelevant as far as observables are concerned.

\subsection{Maximal fake range}
\label{sec:maxfake}

So far, we have presented our results using improved LO and NLO potentials with fake ranges carefully selected to be smaller than the range of the $\rm ^4He$-$\rm ^4He$ interaction. Here, we will describe effects which become increasingly dominant as the fake range increases 
and becomes comparable 
to the size of the atomic interaction. 

\begin{figure*}[t]
\centering
\begin{tabular}{cc}
  \includegraphics[width=0.5\linewidth]{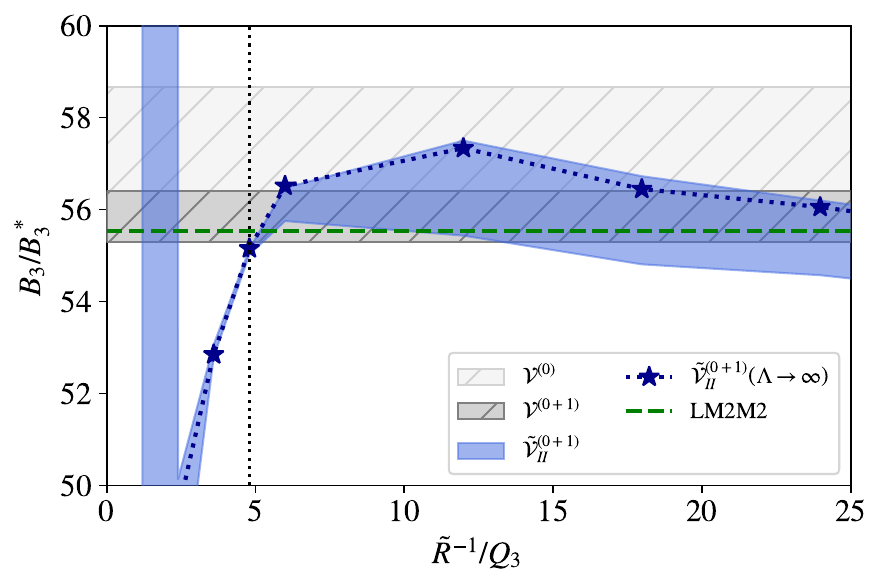}
&
  \includegraphics[width=0.5\linewidth]{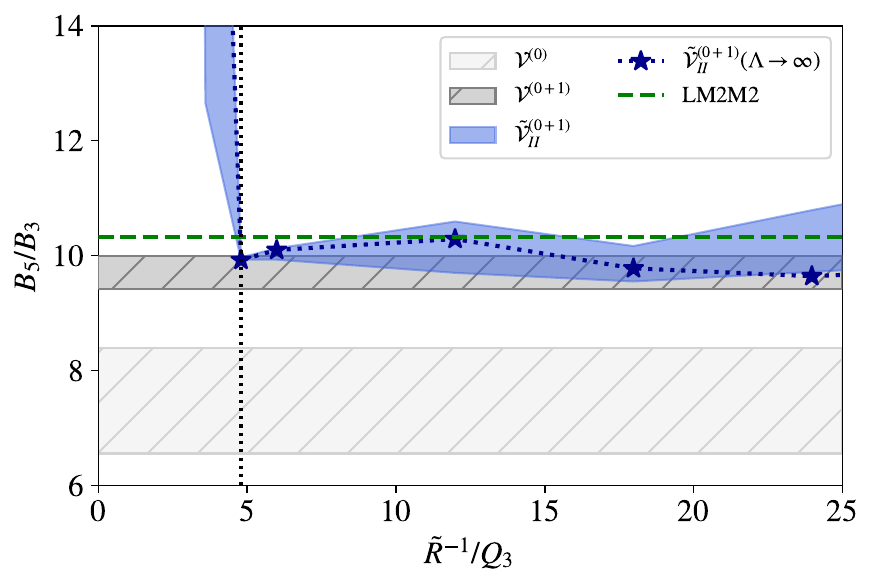}\\
  \caption{Binding energies for the excited trimer $B_3^*$ (left panel) and ground pentamer $B_5$ (right panel) up to NLO, normalized to the ground-trimer binding energy $B_3$, as function of the inverse fake range $\tilde{R}^{-1}$ in units of the three-body scale $Q_3$. 
  Results for the improved NLO potential $\tilde{\mathcal{V}}^{(0+1)}_{II}$
  for cutoff values $\Lambda \ge 6 \, Q_3$ are given by a (blue) shaded band; extrapolated values for $\Lambda \rightarrow \infty$ at the given $\tilde{R}^{-1}$ are marked by (blue) stars. For comparison, the 
  variation of
  LO ($\mathcal{V}^{(0)}$) and NLO ($\mathcal{V}^{(0+1)}$) results
  for cutoff $\Lambda \ge 6 \, Q_3$
  are shown by light and dark (gray) hatched bands, respectively. 
 The (green) horizontal dashed lines mark the result of the full LM2M2 potential \cite{Hiyama:2011ge,doi:10.1063/1.481404}. The vertical dotted line indicates the fake range value, 
  $\tilde{R}_c^{-1} \simeq 4.8\, Q_3$, where the improved LO potential $\tilde{\mathcal{V}}_{II}^{(0)}$ exactly reproduces the LM2M2 effective range.}
  \label{fig:EFTwall}
  \end{tabular}
\end{figure*}

In Fig~\ref{fig:EFTwall}, we display the dependence of our  $\tilde{\mathcal{V}}^{(0+1)}_{II}$ results for the trimer and pentamer on the 
inverse fake range $\tilde{R}^{-1}$. We show the
values for both $B_3/B_3^*(\Lambda\to\infty)$ 
and $B_5(\Lambda\to\infty)/B_3$
obtained from the extrapolation 
with Eq.~(\ref{eq:LO_extrap}) at fixed $\tilde{R}^{-1}$.
We also display the spread of 
results 
for $\Lambda \ge 6Q_3$.
For comparison, we show also the corresponding bands for
the standard interactions $\mathcal{V}^{(0)}$ and $\mathcal{V}^{(0+1)}$.

The figure demonstrates that the extrapolated results become consistent with the standard EFT formulation, $\mathcal{V}^{(0+1)}$, as both the fake range and the cutoff approach the contact limit $\tilde{R}^{-1}\rightarrow \infty$ and $\Lambda \rightarrow \infty$, where $\tilde{\mathcal{V}}^{(0+1)}_{II} = \mathcal{V}^{(0+1)}$. For quite a range of fake-range values, the arbitrary choice of fake-range improvement at LO can be compensated perturbatively at NLO. Both the positions and widths of the NLO bands are similar.

As $\tilde{R}^{-1}$ decreases,
there are two special $\tilde{R}^{-1}$ values visible in Fig.~\ref{fig:EFTwall}.
As we discussed in the previous subsection, 
$\tilde{R}^{-1} = 12 \, Q_3$ represents the situation where our NLO results differ the most from the ``exact'' LM2M2 result, 
likely  
due to the induced subleading correction in the three-body force.
If that is the case, then the difference will be largely suppressed at N$^2$LO, where the first three-body subleading contribution enters. 
The second special value, $\tilde{R}_c^{-1}=4.8\, Q_3$, represents the situation at which the LM2M2 physical effective range $r_2$ is reasonably well reproduced in our calculations already at LO. In this case, NLO corrections besides the four-body scale are 
absorbed in the LO 
fake range,
and the variation of our few-body NLO results with respect to the cutoff $\Lambda$ is minimal. Moreover, at this small $\tilde{R}^{-1}$ the corresponding potentials become reasonably soft, which greatly facilitates their treatment in few- and many-body calculations. Consequently, it might be argued that this is the most convenient point to fix the LO fake range. 

On the other hand, 
it is also a delicate choice. 
Once $\tilde{R}^{-1}< \tilde{R}_c^{-1}$
results begin to reveal increasing fluctuations and 
quickly deteriorate. 
Continuing past this point, the resummation procedure likely becomes nonperturbative, and the standard power counting can no longer be sustained.
This sets a hard limit for the LO resummation in terms of a maximal fake range $\tilde{R}$ that is allowed in order to maintain theory convergence. 
The deterioration of our results is more dramatic for the pentamer. With the limitations discussed above, we find a similar outcome when employing 
$\tilde{\mathcal{V}}^{(0)}_{I}$. 
Since the ``wall'' gets steeper when going from trimer to pentamer, the choice $\tilde{R}=\tilde{R}_c$ might become even more delicate for larger clusters.  

From these results, we conclude that a theory with the modified LO range is equivalent to the standard Contact EFT as long as $\tilde{R}\simle r_2$,
effectively setting the boundary for possible theory resummation.

\section{Discussion and conclusions}
\label{sec:discussion}

We have described how we can improve the action of a renormalizable and convergent power counting by taking advantage of the partial flexibility in the definition of each order.
Higher-order EFT interactions are increasingly singular and 
demand perturbation theory for renormalization.
Model independence 
requires insensitivity to the regularization procedure and its cutoff. Adding to LO subleading effects that preserve order-by-order renormalizability can provide benefits such as improved convergence and easier access to numerical results. 

We based our improvement on the observation that sometimes finite values of the cutoff $\Lambda$, which endows
LO interactions with a range, give results closer to exact results than the $\Lambda\to \infty$ limit. Thus, we used two types of LO improvement which include an arbitrary fake range for the two-body interaction, with one of them also including the same fake range in the three-body force.
We have illustrated this technique employing as benchmark a well-known phenomenological potential for
$\rm ^4He$ atoms. 
Previous studies \cite{Ji:2012nj,Bazak:2016wxm,Bazak:2018qnu} had already demonstrated that an EFT with contact interactions can reproduce the binding energies of systems of up to six atoms systematically. We replicated these results and used them for comparison.
To test our approach further, we conducted our study using 
also a second 
potential model as well as two different forms for NLO interactions
--- non-local and local versions.
We found no significant differences in the results and conclusions 
among these variations.

As could be expected, improved-LO results were almost always better than 
unimproved results.
More crucially, we found that NLO 
results were not significantly affected by these 
modifications as long as the LO interaction remained perturbatively close to the contact limit. In other words, for an extensive range of values the arbitrary choice of fake range is compensated with the perturbative corrections stipulated by the standard power counting. In this range, the improved actions we considered can be used just as other improvements that preserve renormalization-group invariance, such as the choice of low-energy data used to determine interactions strengths.
However, we noticed a 
sudden change in behavior when the arbitrary fake range used at LO increases beyond the physical effective range,
suggesting that using this method 
when they are equal should be done with care.

In spirit, our approach is similar to that of Ref. \cite{Beck:2019abp}, where a finite cutoff was kept at each order, with the hope that the cutoff range could be increased order by order. For the case where we chose to improve both two- and three-body potentials with the same fake range, 
the inverse of our fake range functions as the cutoff at LO in Ref. \cite{Beck:2019abp}. However, our improvement is more general as it does not substitute for a cutoff at LO when the three-body force is not improved, nor for a cutoff at subleading orders.

Our approach also resembles that of Refs. \cite{Kievsky:2020sni,Recchia:2022jih},
where finite-range potentials inspired by EFT are considered. The ranges of the two- and three-body components are varied and a version of the ``wall'' is seen. It is found that the theory error in several few-body ground-state energies is consistent with two-body expectation when the potential ranges are fitted to the two-body effective range and four-body energy. However, an incomplete NLO potential is treated exactly and it is not clear to which extent corrections remain perturbative. If they are not, this approach becomes equivalent to a change in power counting, where a positive effective range is promoted to LO \cite{Beane:2021dab,Timoteo:2023dan}. Our study, where small corrections are treated in perturbation theory, to some extent justifies the results of Refs. \cite{Kievsky:2020sni,Recchia:2022jih}.

The LO improvement is meant to minimize NLO contributions, and the possibility arises that LO interactions can be chosen to effectively reduce the error to N$^2$LO. While the results presented above are promising, in order to achieve full elimination of NLO contributions $\mathcal{V}^{(0)}$ must be improved to account for not only the two-body effective range but also the four-body scale introduced through the four-body force.
It is unlikely that this can be done without including a four-body potential. For the sake of model independence and renormalization, it would be essential to study N$^2$LO contributions to make sure NLO effects have indeed been entirely removed. If successful, this technique could be iterated to resum more subleading aspects of the theory. However, since the number of subleading interactions grows, it will be increasingly difficult to incorporate them in an improved LO.

The more improvements are made, the closer one gets to phenomenological models. Typically such models have a finite range for each component of the potential. These ranges are either fitted to data or varied within narrow intervals (for example, Ref. \cite{Schiavilla:2021dun}). 
One might argue that a phenomenological potential can be used as an auxiliary interaction, essentially transforming a model into the 
lowest order of an improved EFT action. 
A phenomenological potential may include contributions to several orders of a contact theory, partially accounting for them.
However, indiscriminate improvement destroys order-by-order renormalization and obscures power counting. It is crucial to check the renormalizability of the theory and model independence by varying the auxiliary potential used and calculating at least one order of the power counting more than the ones that have been resummed.
Moreover, the calculation of subleading contributions can be simplified by choosing the auxiliary interaction to maximize the stability of numerical codes used in the wave-function calculation. Our improvement is controlled, as it is constrained by the small size of NLO corrections. Because power counting is preserved, the type of interactions and the number of physical parameters at each order are unchanged --- for example, for systems sufficiently closed to unitarity, LO physics is still determined by a single, three-body parameter \cite{Konig:2016utl}. One can also still estimate the theoretical error of a calculation {\it a priori} from the expected size of the first order not accounted for.

The improved-action method might also shed light on why 
additional approximations usually performed in many-body calculations
could
still be compatible with EFTs without necessarily compromising renormalizability. 
Such approximations 
are exemplified by methods like many-body perturbation theory \cite{Tobocman:1957zz} or approaches utilizing the similarity RG
to soften interactions \cite{Glazek:1993rc, Wegner:1994fdg}.
In the former case, only part of the expected LO contributions
is treated exactly~\cite{Drissi:2019lff}, while in the latter induced many-body correlations are typically omitted. 
A possible justification for this procedure is a change in power counting as the number of particles becomes large. (For an example in Chiral EFT, see Ref. \cite{Yang:2021vxa}.)
Or, perhaps, it can be interpreted that these approximations amount to the inclusion of subleading interactions at LO, which cancel the neglected effects. 

We conclude that, even when subleading interactions can be accounted for explicitly, it is beneficial to start from a LO that captures as many of the qualitative aspects of the system as possible. 
For many-fermion systems, in particular, an improved LO might be able to provide a stable state upon which perturbative corrections are much simpler to implement.
We plan to tackle this issue in a future publication.

\vspace{1cm}
\section*{acknowledgement}
We thank N. Barnea, B. Bazak, H.~W.~Grie\ss hammer, A. Lovato, and F. Pederiva for the useful exchanges and discussions.
The work of M. Sch\"{a}fer was supported by the Czech Science Foundation GA\v{C}R grant 22-14497S. This material is based upon work supported in part 
by the U.S. Department of Energy, Office of Science, Office of Nuclear Physics, under award DE-FG02-04ER41338.

\appendix
\section{Local NLO formulation}
\label{App:A}

In addition to the non-local formulation of the NLO presented in Sec. \ref{sec:theory}, Eq.~\eqref{eq:nonlocalNLOprime}; we also employ a simplified, coordinate-space version,
\begin{eqnarray}
    \tilde{V}_{\text{2B}}^{(1)}(\{\vec{r}_i\};\Lambda)
    &=&\tilde{C}_{0}^{(1)}(\Lambda)\sum_{i<j} \delta_\Lambda(\vec{r}_{ij})
    \nonumber
    \\
    &+& \tilde{C}_{2}^{(1)}(\Lambda)\sum_{i<j} r^2_{ij} \delta_\Lambda(\vec{r}_{ij}).
    \label{eq:localNLO}
\end{eqnarray}

These interactions differ from each other by derivatives of the wavefunction, which, in the contact limit, affect only non-zero relative angular momentum of the two particles. 
In this appendix, we derive the local NLO formulations starting from the standard EFT operator of the NLO effective-range term.
In particular, we consider the NLO contribution to the energy of a pair of particles using the Gaussian regulator (Eq. \eqref{eq:regulator}).

First, note that one can write 
\begin{equation}
    \bra{\psi^{(0)}}\left(p'^2+p^2\right)\ket{\psi^{(0)}}=
    \bra{\psi^{(0)}}\left[\left(\vec{p}\,'-\vec{p}\right)^2
    +2\vec{p}\,'\cdot \vec{p}\right]\ket{\psi^{(0)}}.
    \label{eq:approx_local}
\end{equation}
Where $p$ and $p'$ are the incoming and outgoing momenta of the particles in the center of mass frame and $\psi^{(0)}$ is the LO wavefunction of the state on which the perturbation theory is applied.
Next, if we denote the Fourier transform of $\delta_\Lambda(\vec{r})$ by $f_\Lambda(\vec{p})$, then 
\begin{equation}
    f_\Lambda(\vec{p}\,')f_\Lambda(\vec{p})
    =f_\Lambda(\vec{p}-\vec{p}\,')\left[1+{\cal O} \left(\frac{\vec{p}\,'\cdot \vec{p}}{\Lambda^2}\right) \right].
\end{equation}
The $\vec{p}\,'\cdot \vec{p}$ terms contribute only to $P$ and higher waves. Thus, denoting $q=p-p'$ and up to contributions to partial waves higher than $S$, we can write 
\begin{widetext}
\begin{align}
&\bra{\psi^{(0)}}\left(p'^2+ p^2\right)\ket{\psi^{(0)}}
 =\int\frac{d^3p\,'}{(2\pi)^{3/2}} \int\frac{d^3p}{(2\pi)^{3/2}} 
 \, \psi^{*(0)}(\vec{p}\,') \, \psi^{(0)}(\vec{p}) \, f_\Lambda(\vec{p}-\vec{p}\,') \left(\vec{p}\,'-\vec{p}\right)^2 +\ldots
 \nonumber\\
&=\int\frac{d^3r\,'}{(2\pi)^{3/2}}\, \psi^{*(0)}(\vec{r}\,')
 \int\frac{d^3r}{(2\pi)^{3/2}}\, \psi^{(0)}(\vec{r})
 \int\frac{d^3p\,'}{(2\pi)^{3/2}} \, e^{-i\vec{p}\,'\cdot (\vec{r}\,'-\vec{r})}
 \int\frac{d^3q}{(2\pi)^{3/2}} \, e^{i\vec{q}\cdot \vec{r}} \, \vec{q}^{\,2} \, f_\Lambda(\vec{q}) +\ldots 
 \nonumber\\
&=-\int\frac{d^3r'}{(2\pi)^{3/2}}\psi^{*(0)}(\vec{r}\,')
 \int\frac{d^3r}{(2\pi)^{3/2}}\psi^{(0)}(\vec{r})
 \, \delta(\vec{r}\,'-\vec{r}) \, \nabla^2\delta_\Lambda(\vec{r}) +\ldots\\
&=\frac{\Lambda^2}{4} \int\frac{d^3r}{(2\pi)^{3/2}} \, 
\psi^{*(0)}(\vec{r}) \, \psi^{(0)}(\vec{r}) \left(2-r^2\Lambda^2\right) 
\delta_\Lambda(\vec{r}) +\ldots \; . \nonumber
 \end{align}
\end{widetext}
The ``$\ldots$'' contain induced contributions to $P$ waves.
We notice that these contributions are suppressed by only one power of the breakdown scale, while $P$-wave interactions are expected to be suppressed by three powers. However, in $S$ or $D$-wave observables, at least two such interactions are needed, which makes these interactions to induce errors at least one order higher than we work at.
From this analysis, we deduce that the results derived from the derivative and local NLO interactions are equivalent within the framework of this study and its outcomes. 
This conclusion is further substantiated by the data provided in App. \ref{sec:tables}.

\section{Tables of results}
\label{sec:tables}

In this appendix we list the numerical results obtained for the binding energies of the systems considered in this work at various values of 
the fake range
$\tilde{R}$ for both $\tilde{\mathcal{V}}_{I}^{(0)}$ and $\tilde{\mathcal{V}}_{II}^{(0)}$ interactions, Eqs. \eqref{curlyVI} and \eqref{curlyVII}. 
For comparison we also present results for $\tilde{R}=0$ obtained from the ``standard'' $\mathcal{V}^{(0)}$, Eq.~\eqref{VLO}. For the NLO interaction, we consider the two forms, two-derivative, Eq.~\eqref{eq:nonlocalNLOprime}, and $r^2$, Eq.~\eqref{eq:localNLO}. 
The binding energies $B_N^{(*)}$, $N=2-5$, are given at LO and NLO with input parameters fitted to two different sets of low-energy data predicted with the LM2M2 or PCKLJS potential models, see Table~\ref{tab:observables}.
The results for LM2M2 are plotted in the main text.
All results are extrapolated in the regulator cutoff $\Lambda$ according to Eq. \eqref{eq:LO_extrap} and the extrapolation error 
is given.
No EFT truncation errors are shown.

At the two-body level, the interactions $\tilde{\mathcal{V}}_{I}^{(0)}$ and $\tilde{\mathcal{V}}_{II}^{(0)}$ are equivalent.
In Table~\ref{tab:extrapolated_results2} we give our results for the two-body binding energy $B_2$ normalized to the respective model binding energy $B_2^{\rm Model}$. Results are essentially indistinguishable except perhaps for the local NLO potential at the largest value of the fake range for PCKLJS, where it differs by about $3\%$ from the direct model result and the NLO non-local interaction.

\begin{table*}[bt]
    \centering
\begin{tabular}{||c || c c c | c c c ||}
\hline
&\multicolumn{6}{c||}{$B_2/B_2^{\text{Model}}$} \\
& \multicolumn{3}{c|}{LM2M2} & \multicolumn{3}{c||}{PCKLJS} \\
$(Q_3 \tilde{R})^{-1}$ &	LO&	NLO $\nabla^2$&		NLO $r^2$&		LO&	NLO $\nabla^2$&		NLO $r^2$	\\ 
\hline
4.8 & 0.99 & 1.00 & 0.99 & 0.99 & 1.00 & 1.03 \\
6.0 & 0.98 & 1.00 & 1.00 & 0.98 & 1.00 & 1.00 \\
12 & 0.95 & 0.99 & 0.99 & 0.95 & 1.00 & 1.00 \\
18 & 0.94 & 0.99 & 1.00 & 0.94 & 0.99 & 1.00 \\
24 & 0.93 & 0.99 & 1.00 & 0.93 & 0.99 & 1.00 \\
30 & 0.93 & 0.99 & 1.00 & 0.93 & 0.99 & 1.00 \\
36 & 0.93 & 0.99 & 1.00 & 0.93 & 0.99 & 1.00 \\
42 & 0.93 & 0.99 & 1.00 & 0.93 & 0.99 & 1.00 \\
$\infty$ & 0.93 & 0.99 & 1.00 & 0.93 & 0.99 & 1.00 \\\hline
\end{tabular}
    \caption{Cutoff-extrapolated ($\Lambda\rightarrow\infty$) dimer binding energy $B_2$
    for several values of the inverse fake range $\tilde{R}^{-1}$, in units of the three-body scale $Q_3$.
    The binding energy was calculated with the LO potential $\tilde{\mathcal{V}}_{I}^{(0)}$ (equivalent at two-body level to $\tilde{\mathcal{V}}_{II}^{(0)}$)
    and two versions of the corresponding NLO interaction, 
    $\nabla^2$ and $r^2$.
    The symbol ``$\infty$'' marks the result for 
    the ``standard'' $\mathcal{V}^{(0)}$. 
    Calculated 
    energies are given as ratios with respect to the 
    ``exact'' 
    value $B_2^{\text{Model}}$ listed in Table~\ref{tab:observables} for either LM2M2 or PCKLJS phenomenological models. 
    Extrapolation errors are negligible. No EFT truncation errors are shown.}
    \label{tab:extrapolated_results2}
\end{table*}

Results for $B_{N\ge 3}^{(*)}$ differ slightly depending on whether one employs $\tilde{\mathcal{V}}_{I}^{(0)}$ or $\tilde{\mathcal{V}}_{II}^{(0)}$.
In Tables \ref{tab:extrapolated_results_2range}, \ref{tab:extrapolated_results34}, and \ref{tab:extrapolated_results5}
one finds the binding energies of the excited trimer $B_3^*$ and of the ground tetramer and pentamer, $B_4$ and $B_5$ respectively. They are normalized to the ground trimer energy $B_3$, which is fitted to the respective potential-model result.

\begin{table*}[bt]
    \centering
\begin{tabular}{||c|| c | c|| c| c|| c| c||}
\hline
&\multicolumn{2}{c||}{$B_3/B_3^*$}&\multicolumn{2}{c||}{$B_4/B_3$}&\multicolumn{2}{c||}{$B_5/B_3$}\\
$(Q_3 \tilde{R})^{-1}$ &LM2M2&PCKLJS&LM2M2&PCKLJS&LM2M2&PCKLJS\\ \hline
4.8 & 55.656	&	49.688	& 4.36(0)&    4.30(0) & 10.1(1)	&	9.9(3)\\
6.0 & 56.986	&	50.947	& 4.12(1)&	4.07(1) & 9.2(1)&	9.0(1)\\
12 & 58.068	&	52.100	& 3.98(5)&	3.93(5) & 8.6(6)&	8.5(5)\\
18 & 58.418	&	52.469	& 4.01(6)&	3.96(6) & 8.6(3)&	8.5(1)\\
24 & 58.661	&	52.728	& 4.04(4)&	4.00(4) & 8.63(1)& 8.5(3)\\
30 & 58.762	&	52.868	& 4.08(2)&	4.04(2) & 8.7(4)$^*$&	8(1)$^*$\\
36 & 58.981	&	53.061	& 4.12(3)&	4.07(3) & 8.8(9)$^*$&	9(2)$^*$\\
42 & 59.204    &   53.277 & 4.15(1)&    4.11(1) & 9(2)$^*$&	8(3)$^*$\\
$\infty$ & 59.250    &   53.335 & 4.14(1)&    4.10(1)& 9.27(5) &9.13(5) \\\hline
Model & 55.482    &   49.751 & 4.42 & 4.35 & 10.33 & -\\\hline
\end{tabular}
    \caption{Cutoff-extrapolated ($\Lambda\rightarrow\infty$) LO binding energies $B_3^*$ of the trimer excited state, $B_4$ of the tetramer ground state, and $B_5$ of the pentamer ground state
    for several values of the inverse fake range $\tilde{R}^{-1}$, in units of the three-body scale $Q_3$. 
    Binding energies were calculated with the LO potential $\tilde{\mathcal{V}}_{I}^{(0)}$. They are given as ratios with respect to the ``exact'' trimer ground energy $B_3$ listed in Table~\ref{tab:observables} for either LM2M2 or PCKLJS phenomenological models, which is fitted in the EFT.
    Extrapolation errors, if sizeable, are given in parenthesis. 
    The pentamer values marked with an asterisk
    are suspected to be less numerically stable than 
    other results due to the large $\tilde{R}^{-1}$ employed; although this is reflected in the large extrapolation errors, they should be taken with caution.
    The last row gives the corresponding ratios calculated directly with LM2M2 and PCKLJS potential models \cite{LM2M2:1991,PCKLJS:2010}.}
    \label{tab:extrapolated_results_2range}
\end{table*}

\begin{table*}[bt]
    \centering
\begin{tabular}{||c || c c c | c c c || c c | c c ||}
\hline
&\multicolumn{6}{c||}{$B_3/B_3^*$} & \multicolumn{4}{c||}{$B_4/B_3$}\\
& \multicolumn{3}{c|}{LM2M2} & \multicolumn{3}{c||}{PCKLJS} & \multicolumn{2}{c|}{LM2M2} & \multicolumn{2}{c||}{PCKLJS} \\
$(Q_3 \tilde{R})^{-1}$ &	LO&	NLO $\nabla^2$&		NLO $r^2$&		LO&	NLO $\nabla^2$&		NLO $r^2$	& LO& NLO 
& LO & NLO 
\\ \hline
4.8&	55.024	& 55.121		& 55.024		& 49.176	& 49.232		& 49.241 &	- & \parbox[t]{2mm}{\multirow{6}{*}{\rotatebox[origin=c]{90}{Fitted}}}  & -&\parbox[t]{2mm}{\multirow{6}{*}{\rotatebox[origin=c]{90}{Fitted}}} \\
6.0&	55.949	& 56.479		& 56.479		& 50.091	& 50.486		& 50.473 &	3.76& &3.71& \\
12&	57.605	& 58.112		& 58.060		& 51.722	& 51.869		& 51.828 &	3.41& &3.36& \\
18&	58.107	& 57.234		& 57.087		& 52.235	& 51.229		& 51.022 &	3.58&	&3.54&\\
24&	58.565	& 56.523		& 56.173		& 52.589	& 50.533		& 50.193 &	3.72&	&3.68&\\
$\infty$&	59.089	& 55.803 & 55.964 & 52.729 & 49.833 & 49.995 &	4.14(1)& &4.10(1)& \\\hline
Model & \multicolumn{3}{c|}{55.533}&\multicolumn{3}{c||}{49.747}& \multicolumn{2}{c|}{4.42}&\multicolumn{2}{c||}{4.35}\\\hline
\end{tabular}
\caption{
Cutoff-extrapolated ($\Lambda\rightarrow\infty$) LO and NLO binding energies $B_3^*$ of the trimer excited state and $B_4$ of the tetramer ground state
    for several values of the inverse fake range $\tilde{R}^{-1}$, in units of the three-body scale $Q_3$. Binding energies were calculated with the LO potential $\tilde{\mathcal{V}}_{II}^{(0)}$
    and two versions of the corresponding NLO interaction, 
    $\nabla^2$ and $r^2$. 
    They are given as ratios with respect to the ``exact'' trimer ground energy $B_3$ listed in Table~\ref{tab:observables} for either LM2M2 or PCKLJS phenomenological models, which is fitted in the EFT.
    Extrapolation errors, if sizeable, are given in parenthesis. 
    The tetramer $B_4/B_3$ ratio has been fitted to adjust the strength of the NLO four-body force. 
    The last row gives the corresponding ratios calculated directly with LM2M2 and PCKLJS potential models \cite{LM2M2:1991,PCKLJS:2010}.}
    \label{tab:extrapolated_results34}
\end{table*}

\begin{table*}[bt]
    \centering
\begin{tabular}{||c || c c c | c c c ||}
\hline
&\multicolumn{6}{c||}{$B_5/B_3$}\\
& \multicolumn{3}{c|}{LM2M2} & \multicolumn{3}{c||}{PCKLJS} \\
$(Q_3 \tilde{R})^{-1}$ &	LO&	NLO $\nabla^2$&		NLO $r^2$&		LO&	NLO $\nabla^2$&		NLO $r^2$		\\ \hline
4.8 & 9.68 & 9.92  (0) & 10.25(0) & 9.50 & 9.74(0) & 9.85(0) \\
6.0 & 7.40 & 10.10(1) & 10.13(1) & 7.28 & 9.89(1) & 10.41(5) \\
12 & 6.56 & 10.29(5) & 10.60(5) & 6.44 & 10.09(5) & 9.88(6) \\
18 & 7.21 & 9.78(6) & 10.07(6) & 7.09 & 9.59(6) & 9.62(4) \\
24 & 7.70 & 9.64(4) & 9.82(4) & 7.57 & 9.44(4) & 9.66(2) \\
$\infty$&9.27(5)	& 9.76(1)	& 10.05(1)&	9.13(5)	& 9.57(1)	& 9.86(1)	\\\hline
Model & \multicolumn{3}{c|}{10.33}&\multicolumn{3}{c||}{-}\\\hline
\hline
\end{tabular}
    \caption{Same as in Table~\ref{tab:extrapolated_results2} but for the pentamer binding energy $B_5$.
    }
    \label{tab:extrapolated_results5}
\end{table*}

LO binding-energy ratios obtained with $\tilde{\mathcal{V}}_{I}^{(0)}$ are reported in Table \ref{tab:extrapolated_results_2range}. We present a large range of fake ranges.
For the smaller values of $\tilde{R}$, pentamer results are probably less reliable than the rest. For comparison, the corresponding ratios calculated directly with LM2M2 and PCKLJS potential models \cite{LM2M2:1991,PCKLJS:2010} are also given. As it can be seen, the trend for PCKLJS results is similar to LM2M2: results are quite close to the ``exact'' results at the largest value of the fake range and end up about 10\% off for $\tilde{R}=0$.   

Results obtained with $\tilde{\mathcal{V}}_{II}^{(0)}$ at LO and the two forms of the NLO interaction --- non-local and local --- are reported in Tables~\ref{tab:extrapolated_results34} and \ref{tab:extrapolated_results5}. LO binding-energy ratios follow the same trend as $\tilde{\mathcal{V}}_{I}^{(0)}$, but are smaller, the difference increasing with particle number and reaching over 10\% for $B_5/B_3$. Agreement with ``exact'' results slightly improves for the excited trimer but deteriorates for the tetramer and pentamer. At NLO, the two 
types of potential give the same results within a few percent, and binding energies are essentially independent of the fake range. As expected, the inclusion of the two-body effective range makes a very small effect for the excited trimer when the fake range is numerically close to the experimental  value of the effective range, but as the fake range is removed the effect remains no larger than 10\%. In the end, excellent agreement with the ``exact'' result is obtained. The four-body force is fitted to the tetramer ground-state energy and the combined NLO effects increase the pentamer binding energy. For LM2M2, where an ``exact'' result exists \cite{LM2M2:1991}, NLO halves the pentamer discrepancy. 

We see that the results, although different in detail, are qualitatively the same for LM2M2 and PCKLJS, and for the two versions of the NLO potential. Differences are well within the expected truncation errors discussed in the main text. Results presented in the main text for LM2M2 for the non-local NLO form are therefore representative of the theory.

\bibliography{Bib.bib}
\end{document}